# Cloud Computing

Shivaji P. Mirashe, Dr. N.V. Kalyankar.

**Abstract**— Computing as you know it is about to change, your applications and documents are going to move from the desktop into the cloud.

I'm talking about cloud computing, where applications and files are hosted on a "cloud" consisting of thousands of computers and servers, all linked together and accessible via the Internet. With cloud computing, everything you do is now web based instead of being desktop based. You can access all your programs and documents from any computer that's connected to the Internet.

How will cloud computing change the way you work? For one thing, you're no longer tied to a single computer. You can take your work anywhere because it's always accessible via the web. In addition, cloud computing facilitates group collaboration, as all group members can access the same programs and documents from wherever they happen to be located. Cloud computing might sound far-fetched, but chances are you're already using some cloud applications. If you're using a web-based email program, such as Gmail or Hotmail, you're computing in the cloud. If you're using a web-based application such as Google Calendar or Apple Mobile Me, you're computing in the cloud. If you're using a file- or photo-sharing site, such as Flickr or Picasa Web Albums, you're computing in the cloud. It's the technology of the future, available to use today.

**Index Terms**—What is Cloud Computing?, Cloud computing is Programmable, Understanding Cloud Architecture, Why Computing Advantage and Disadvantage, Who Benefits from Cloud Computing, Cloud Computing for Everyone, Using cloud computing Services, Computing on the Cloud & Privacy, Security, and standards Compliance.

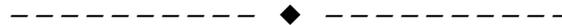

## 1 INTRODUCTION

First, cloud computing isn't network computing. With network computing, application or documents are hosted on a single company's server and accessed over the company's network. Cloud computing is a lot bigger than that. It encompasses multiple companies, multiple servers, and multiple networks. Plus, unlike network computing, cloud services and storage are accessible from anywhere in the world over an Internet connection; with network computing, access is over the company's network only.

Cloud computing also isn't traditional outsourcing, where a company farms out (subcontracts) its computing services to an outside firm. While an outsourcing firm might host a company's data or applications, those documents and programs are only accessible to the company's employees via the company's network, not to the entire world via the Internet. So, despite superficial similarities, networking computing and outsourcing are not cloud computing.

## 2 WHAT IS CLOUD COMPUTING?

How does cloud computing work? What does cloud computing mean for the way you use a computer? What are the top cloud computing applications? Good questions all, and all answered in this paper, Cloud Computing. That Change the Way You Work and Collaborate

- *F.A. is Working as a Manager,IDC-Internet Data Center,Reliance Communications,New Mumbai.(Maharashtra) - (INDIA)*
- *S.A. Principal of Yeshwant Mahavidyalaya Nanded.*

Online. I don't pretend to answer every question you may have (the overly technical ones in particular), but I do try to give you a good solid overview of the cloud computing phenomenon, and introduce you to some of the more popular cloud applications—in particular, those that facilitate group collaboration. And that's where cloud computing really shines. Whether you want to share photographs with family members, coordinate volunteers for a community organization, or manage a multifaceted project in a large organization, cloud computing can help you collaborate and communicate with other group members. You'll have a better idea of how this works after you read the book, but trust me on this one—if you need to collaborate, cloud computing is the Sway to do it. Key to the definition of cloud computing is the "cloud" itself. For our purposes, the cloud is a large group of interconnected computers. These computers can be personal computers or network servers; they can be public or private. For example, Google hosts a cloud that consists of both smallish PCs and larger servers. Google's cloud is a private one (that is, Google owns it) that is publicly accessible (by Google's users). This cloud of computers extends beyond a single company or enterprise. The applications and data served by the cloud are available to broad group of users, cross-enterprise and cross-platform. Access is via the Internet. Any authorized user can access these docs and apps from any computer over any Internet connection. And, to the user, the technology and infrastructure behind the cloud is invisible. It isn't apparent (and, in most cases doesn't matter) whether cloud services are based on HTTP, HTML, XML, JavaScript, or other specific technologies. It might help to examine how one of the pioneers of cloud computing, Google, perceives the topic. From Google's perspective, there are six key properties of cloud computing as below,

**1) Cloud computing is user-centric.** Once you as a user are connected to the cloud, whatever is stored there—documents, messages, images, applications, whatever—becomes yours. In addition, not only is the data yours, but you can also share it with others. In effect, any device that accesses your data in the cloud also becomes yours.

**2) Cloud computing is task-centric.** Instead of focusing on the application and what it can do, the focus is on what you need done and how the application can do it for you., Traditional applications—word processing, spreadsheets, email, and so on—are becoming less important than the documents they create.

**3) Cloud computing is powerful**. Connecting hundreds or thousands of computers together in a cloud creates a wealth of computing power impossible with a single desktop PC. Cloud computing is accessible. Because data is stored in the cloud, users can instantly retrieve more information from multiple repositories. You're not limited to a single source of data, as you are with a desktop PC.

**4) Cloud Computing is intelligent.** With all the various data stored on the computers in a cloud, data mining and analysis are necessary to access that information in an intelligent manner.

**5) Cloud Computing is Programmable.** Many of the tasks necessary with cloud computing must be automated. For example, to protect the integrity of the data, information stored on a single computer in the cloud must be replicated on other computers in the cloud. If that one computer goes offline, the cloud's programming automatically redistributes that computer's data to a new computer in the cloud.

All these definitions behind us, what constitutes cloud computing in the real world? Internet-accessible, group-collaborative applications are currently available, with many more on the way. Perhaps the best and most popular examples of cloud computing applications today are the Google family of applications—Google Docs & Spreadsheets, Google Calendar, Gmail, Picasa, and the like. All of these applications are hosted on Google's servers, are accessible to any user with an Internet connection, and can be used for group collaboration from anywhere in the world. In short, cloud computing enables a shift from the computer to the user, from applications to tasks, and from isolated data to data that can be accessed from anywhere and shared with anyone. The user no longer has to take on the task of data management; he doesn't even have to remember where the data is. All that matters is that the data is in the cloud, and thus immediately available to that user and to other authorized users.

## 3 UNDERSTANDING CLOUD ARCHITECTURE ECTIONS

Cloud architecture, the systems architecture of the software systems involved in the delivery of cloud computing, comprises hardware and software designed by a cloud architect who typically works for a cloud integrator. It typically involves multiple cloud components communicating with each other over application programming interfaces, usually web services.

This closely resembles the Unix philosophy of having multiple programs each doing one thing well and working together over universal interfaces. Complexity is controlled and the resulting systems are more manageable than their monolithic counterparts.

Cloud architecture extends to the client, where web browsers and/or software applications access cloud applications.

Cloud storage architecture is loosely coupled, often assiduously avoiding the use of centralized metadata servers which can become bottlenecks. This enables the data nodes to scale into the hundreds, each independently delivering data to applications or users. demonstrated in this document, the numbering for sections upper case Arabic numerals, then upper case Arabic numerals, separated by periods. Initial paragraphs after the section title are not indented. Only the initial, introductory paragraph has a drop cap.

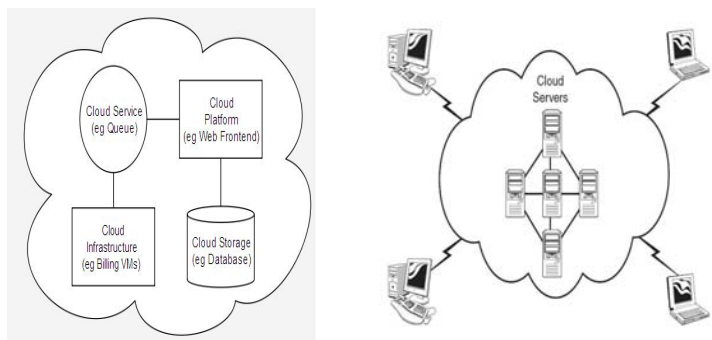

Figure 1 Cloud computing sample architecture. Figure 2 Understanding Cloud Architecture

The key to cloud computing is the "cloud"—a massive network of servers or even individual PCs interconnected in a grid. These computers run in parallel, combining the resources of each to generate supercomputing-like power. What, exactly, is the "cloud"? Put simply, the cloud is a collection of computers and servers that are publicly accessible via the Internet. This hardware is typically owned and operated by a third party on a consolidated basis in one or more data center locations. The machines can run any combination of operating systems; it's the processing power of the machines that matter, not what their desktops look like.

As shown in Figure 1.1, individual users connect to the cloud from their own personal computers or portable devices, over the Internet. To these individual users, the cloud is seen as a single application, device, or document. The hardware in the cloud (and the operating system that manages the hardware connections) is invisible.

## 4 WHY COMPUTING ADVANTAGE AND DISADVANTAGE:-

Any serious analysis of cloud computing must address the advantages and disadvantages offered by this burge-oning technology. What's good—and what's bad—about cloud computing? Let's take a look. **Advantage as below**, We'll start with the advantages offered by cloud computing—and there are many.
1) Lower-Cost Computers for Users
2) Improved Performance
3) Lower IT Infrastructure Costs
4) Fewer Maintenance Issues
5) Lower Software Costs
6) Instant Software Updates
7) Increased Computing Power
8) Unlimited Storage Capacity
9) Increased Data Safety
10) Improved Compatibility Between Operating Systems
11) Improved Document Format Compatibility
12) Easier Group Collaboration
13) Universal Access to Documents
14) Latest Version Availability
15) Removes the Tether to Specific Devices
   **Disadvantage as below,**
That's not to say, of course, that cloud computing is without its disadvantages. There are a number of reasons why you might not want to adopt cloud computing for your particular needs. Let's examine a few of the risks related to cloud computing.
1) Requires a Constant Internet Connection
2) Doesn't Work Well with Low-Speed Connections
3) Can Be Slow
4) Features Might Be Limited
5) Stored Data Might Not Be Secure
6) If the Cloud Loses Your Data, You're screwed.

## 5  WHO BENEFITS FROM CLOUD COMPUTING?

Let's face it, cloud computing isn't for everyone. What types of users, then, are best suited for cloud computing—and which aren't?
   1) Collaborators
   2) Road Warriors
   3) Cost-Conscious Users
   4) Cost-Conscious IT Departments
   5) Users with Increasing Needs
**Collaborators:-**
   If you often collaborate with others on group projects, you're an ideal candidate for cloud computing. The ability to share and edit documents in real time between multiple users is one of the primary benefits of web-based applications. it makes collaborating easy and even fun.
**Road Warriors:-**
    Another prime candidate for cloud computing is the road warrior. When you work at one office today, at home the next day, and in another city the next, it's tough to keep track of all your documents and applications. You may end up with one version of a document on your work PC, another on your laptop, and a third on your home PC—and that's if you remember to copy that document and take it with you from one location to the next.

## 6  CLOUD COMPUTING FOR EVERYONE?

Now that you know a little bit about how cloud computing works, let's look at how you can make cloud computing work for you. By that I mean real-world examples of how typical users can take advantage of the collaborative features inherent in web-based applications.
       We'll start our real-world tour of cloud computing by examining how an average family can use web-based applications for various purposes. As you'll see, computing in the cloud can help a family communicate and collaborate—and bring family members closer together.
**I-Cloud Computing for the Family**
    1) Centralizing Email Communications
    2) Collaborating on Schedules
    3) Collaborating on Grocery Lists
    4) Collaborating on To-Do Lists
    5) Collaborating on Household Budgets
    6) Collaborating on Contact Lists
    7) Collaborating on School Projects
    8) Sharing Family Photos
**II-Cloud Computing for the Community**
    1) Communicating Across the Community
    2) Collaborating on Schedules
    3) Collaborating on Group Projects and Events
**III- Cloud Computing for the Corporation**
    1) Managing Schedules
    2) Managing Contact Lists
    3) Managing Project
    4) Collaborating on Reports
    5) Collaborating on Marketing Materials
    6) Collaborating on Expense Reports.
    7) Collaborating on Budgets
    8) Collaborating on Financial Statements
    9) Collaborating on Presentation

## 7  USING CLOUD COMPUTING SERVICE:-

Services provided by cloud computing can be split into three major categories.
**i)Infrastructure-as-a-Service (IaaS):-**
Infrastructure-as-a-Service (IaaS) like Amazon Web Services provides virtual servers with unique IP addresses and blocks of storage on demand. Customers benefit from an API from which they can control their servers. Because customers can pay for exactly the amount of service they use, like for electricity or water, this service is also called utility computing.
  **ii) Software-as-a-Service (SaaS):-**
Software-as-a-Service (SaaS) is the broadest market. In this case the provider allows the customer only to use its applications. The software interacts with the user through a user interface. These applications can be anything from web based email, to applications like Twitter or Last.fm.

## 7  COMPUTING ON THE CLOUD:-

Cloud computing is offered in different forms as below,
1) Public Cloud.
2) Private Cloud

3) Hybrid cloud which combine both Public & Private.

**Public Cloud:-**

Public cloud or external cloud describes cloud computing in the traditional mainstream sense, whereby resources are dynamically provisioned on a fine-grained, self-service basis over the Internet, via web applications/web services, from an off-site third-party provider who shares resources and bills on a fine-grained utility computing basis.

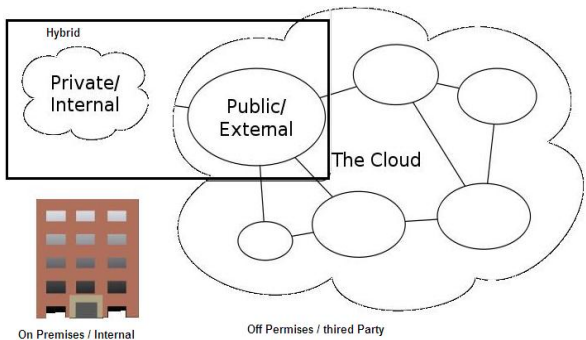

Figure of Cloud computing types

**Hybrid cloud:-**

A hybrid cloud environment consisting of multiple internal and/or external providers "will be typical for most enterprises". A hybrid cloud can describe configuration combining a local device, such as a Plug computer with cloud services. It can also describe configurations combining virtual and physical, colocated assets—for example, a mostly virtualized environment that requires physical servers, routers, or other hardware such as a network appliance acting as a firewall or spam filter.

**Private cloud:-**

Private cloud and internal cloud are neologisms that some vendors have recently used to describe offerings that emulate cloud computing on private networks. These (typically virtualisation automation) products claim to "deliver some benefits of cloud computing without the pitfalls", capitalising on data security, corporate governance, and reliability concerns. They have been criticized on the basis that users "still have to buy, build, and manage them" and as such do not benefit from lower up-front capital costs and less hands-on management, essentially "[lacking] the economic model that makes cloud computing such an intriguing concept".

While an analyst predicted in 2008 that private cloud networks would be the future of corporate IT, there is some uncertainty whether they are a reality even within the same firm. Analysts also claim that within five years a "huge percentage" of small and medium enterprises will get most of their computing resources from external cloud computing providers as they "will not have economies of scale to make it worth staying in the IT business" or be able to afford private clouds. Analysts have reported on Platform's view that private clouds are a stepping stone to external clouds, particularly for the financial services, and that future datacenters will look like internal clouds.

The term has also been used in the logical rather than physical sense, for example in reference to platform as a service offerings, though such offerings including Microsoft's Azure Services Platform are not available for on-premises deployment

## 9 PRIVACY, SECURITY, AND STANDARDS COMPLIANCE:-

A major issue in cloud computing, especially with public clouds, is protection of user data. One concern is that cloud providers themselves may have access to customers' unencrypted data - whether it's on disk, in memory, or transmitted over the network. To limit this exposure, many sources recommend never giving providers access to unencrypted data or keys. A second concern is that many public cloud providers are unable or unwilling to allow auditing of their physical or network security measures. This can preclude them, and thus their customers, from meeting standards such as the US government's HIPAA or Sarbanes-Oxley, the European Union's Data Protection Directive, or the credit card industry's PCI DSS. The extent of some public clouds across multiple legal jurisdictions further complicates this issue; see "Legal Issues" for more detail. These concerns are considered key obstacles to broader adoption of cloud computing, making them areas of active research and debate among cloud computing practitioners and advocates.

## 10 CONCLUSION: -

We are observes that cloud computing has been defined as "everything that we currently do". Many technologies that have been branded as "cloud computing" have existed for a long time before the "cloud" label came into existence. Examples include databases, load balanced on-demand web hosting services, network storage, real time online services, hosted services in general.

## 11 ACKNOWLEDGMENT :-

The authors wish to thank Miss. Suvran D. Alandkar (Mirash), Professor Shivaji Balaji Chavan (Yeshwant Mahavidyalaya Nanded), Mr. Aniket Despande (IBM), Mr. Satish Khadap (infovisionindia Consulting Services) & Mr.Satish D. Alandkar (BSNL-Pune). This work was supported in part by a grant from support to write this paper.

## 12 REFERENCES:-

**First Author:- Shivaji P. Mirashe**

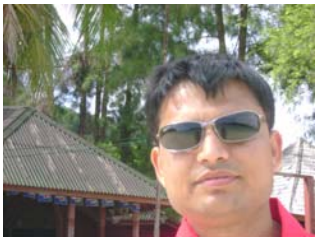

**Myself Mr. Shivaji Pandurangrao Mirashe**. I have completed MCA from S.R.T.M.U. Nanded (Maharashtra) Indian. I have got 8 Years experience in Information Technology. Currently I am working as a Manager in Information Security in Reliance Communication. Mumbai – Maharashtra (INDIA), doing the PHD at Yeshwant Mahavidyalaya Nanded, affiliated to S.R.T.M. University NANDED, Maharashtra, INDIA

Paper Publish at different place & IEEE as below name,

1) Firewall Penetration Testing Paper ID M575 – The 2nd International Conference on Computer modeling and simulation (iccms 2010)- http://iccms.org/
2) Peer-to-Peer Network Protocols ID H271 - The 2010 International Conference on Signal Acquisition and Processing (ICSAP 2010) - http://www.icsap.org/
3) Why We Need the Intrusion Detection Prevention Systems (IDPS) In IT Company – ID E447. - 2nd International Conference on Computer Engineering and Applications (ICCEA 2010) - http://www.iccea.org/
4) E-marketing, Unsolicited Commercial E-mail, and Legal Solutions – Emerging Trends in Computer Science, Communication & Information Technology (CSCIT2010) www.cscit2010.com
5) Saving the World Unsolicited Email Flow - Emerging Trends in Computer Science and Information Technology-2010 (For further information visit http://www.kkwagh.org/ETCSIT/ETCSIT10.html
6) Shivji Mirashe is a member of the IEEE and the IEEE Computer Society & International Association of Computer Science and Information Technology
IACSIT( Member NO. : 80337345) .

**Second Author:-**
**Namdeo V. Kalyankar:**
Dr. N.V. Kalyankar
Principal of Yeshwant Mahavidyalaya Nanded.
S.R.T.M.University
Nanded (Maharashtra) - (INDIA).
Completed M.Sc. Physics from B.A.M. University
, Aurangabad. in 1980. in 1980 he joined as Lecturer in Department of Physics in yeshwant College,Nanded. In 1984 he completed his DHE. He Completed his Ph.D. from B.A.M.University in 1995. From 2003 he is working as Principal since 2003 to till date in Yeshwant college Nanded. He is also Research Guide for Computer Studies in S.R.T.M. University , Nanded. He is also worked on various bodies in S.R.T.M. University Nanded. He also published research papers in various international/ national journals. He is peer team member of NAAC (National Assessment and Accreditation Council)(India). He published a book entitled " DBMS Concept and programming in Foxpro". He also got "Best Principal" award from S.R.T.M. University, Nanded(India) in 2009. He is life member of Indian National Congress , Kolkata (India). He is also honored with "Fellowship of Linnean Society of London (F.L.S.)" on 11th Nov. 2009.